\documentclass[preprint,12pt,3p]{elsarticle}

\usepackage{amsmath,amssymb,amsfonts}
\usepackage{mathrsfs}
\usepackage{chemformula}
\usepackage{color}
\usepackage[normalem]{ulem}
\usepackage{stmaryrd}
\usepackage{diagbox}
\usepackage[parfill]{parskip}

 \biboptions{sort&compress}

\newcommand{\blue}[1]{\textcolor{black}{#1}}

\newcommand{\pfr}[2]{\ensuremath{\frac{\partial #1}{\partial #2}}}
\newcommand{\pfi}[2]{\ensuremath{{\partial #1}/{\partial #2}}}
\newcommand{\ep}{\epsilon}
\newcommand\Lew{\mbox{\textit{Le}}}
\newcommand\Pra{\mbox{\textit{Pr}}}
\newcommand\Pec{\mbox{\textit{Pe}}}
\newcommand\Dam{\mbox{\textit{Da}}}
\newcommand\Ray{\mbox{\textit{Ra}}}
\newcommand{\vect}[1]{\mathbf{#1}}

\journal{Proceedings of the Combustion Institute}

\begin{document}

\begin{frontmatter}

\title{Effect  of a shear flow on the Darrieus--Landau instability in a Hele-Shaw channel}

\author{Prabakaran Rajamanickam, Joel Daou}
\address{Department of Mathematics, University of Manchester, Manchester M13 9PL, UK}

\begin{abstract}
 The Darrieus--Landau instability of premixed flames propagating in a narrow Hele-Shaw channel in the presence of a strong shear flow is investigated, incorporating also  the Rayleigh--Taylor  and diffusive-thermal instabilities. The flow induces shear-enhanced diffusion (Taylor dispersion)  in the two-dimensional depth averaged   equations. Since the diffusion enhancement is in the streamwise direction, but not in the spanwise direction, this leads to anisotropic diffusion and  flame propagation. To understand how such anisotropies  affect   flame  stability, two  important cases are considered. These correspond to initial unperturbed  conditions pertaining to a planar flame propagating  in the streamwise or  spanwise directions. The  analysis is based on a  two-dimensional model derived by asymptotic methods and solved numerically. Its numerical solutions comprise the computation of eigenvalues of a linear stability problem as well as time-dependent simulations. These address the influence of the shear-flow strength (or  Peclet number $\Pec$), preferential diffusion (or Lewis number $\Lew$) and gravity (or Rayleigh number $\Ray$). Dispersion curves characterizing the perturbation  growth rate  are computed for selected values of $\Pec$, $\Lew$ and $\Ray$.  Taylor dispersion induced by strong shear flows is found to suppress the Darrieus--Landau  instability and to weaken the flame wrinkling when the flame propagates in the streamwise direction. In contrast, when the flame propagates in the spanwise direction,  the flame is stabilized in $\Lew<1$ mixtures, but destabilized in $\Lew>1$ mixtures. In the latter case, Taylor dispersion coupled with gas expansion  facilitates  flame wrinkling in an unusual manner.  Specifically, stagnation points and counter-rotating vortices are encountered in the flame close to the unburnt gas side. More generally, an original finding is the demonstration that vorticity  can be produced by a curved flame in a Hele-Shaw channel even in the absence of gravity, whenever $\Pec \neq 0$, and that the vorticity remains confined to the flame preheat and reaction zones. 
\end{abstract}

\begin{keyword}
    Darrieus--Landau instability \sep Rayleigh--Taylor instability \sep diffusive-thermal instability \sep Taylor dispersion \sep Hele-Shaw burner
\end{keyword}

\end{frontmatter}

\section{Introduction\label{sec:introduction}} 

Several recent experimental and theoretical studies have reported  interesting results on flame dynamics arising due to flame instabilities in Hele-Shaw burners. These include  low Lewis-number dendritic flames~\cite{veiga2020unexpected,gu2021propagation,yanez2022velocity}, the Darrieus--Landau instability~\cite{al2019darrieus,veiga2019experimental,tayyab2020experimental,han2021effect}, slowly-drifting isolated or paired flame rings~\cite{dominguez2023stable}, and  the Rayleigh--Taylor  instability~\cite{fernandez2019impact}. 

One of the  practical and scientifically interesting factors to consider is the effect
of an imposed shear flow in Hele-Shaw channels   on flame propagation and instabilities. We have addressed this problem in a series of recent publications which were specifically dedicated to the flame diffusive-thermal instability in narrow channels within the constant-density approximation~\cite{daou2021effect,daou2023flame,daou2023diffusive}. The main objective of this paper is to extend such studies to investigate  the Darrieus--Landau  instability   in narrow channels, while also accounting for its coupling with the Rayleigh--Taylor and the diffusive-thermal instabilities.

\begin{figure}
\centering
\text{   (a) Streamwise flame propagation}\par\vspace{0.1cm}
\includegraphics[width=0.47\textwidth]{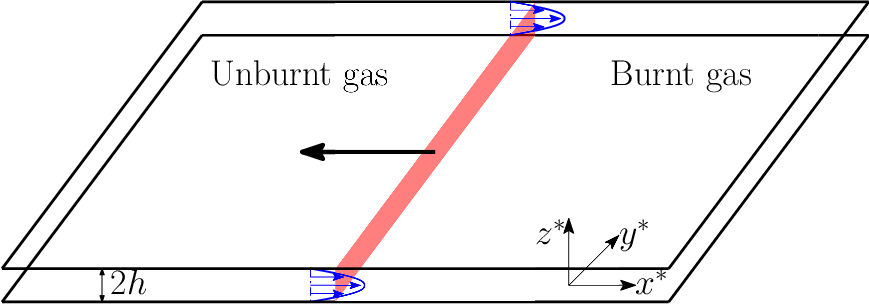}  
\text{ (b) Spanwise flame propagation}\par\vspace{0.1cm}
\includegraphics[width=0.47\textwidth]{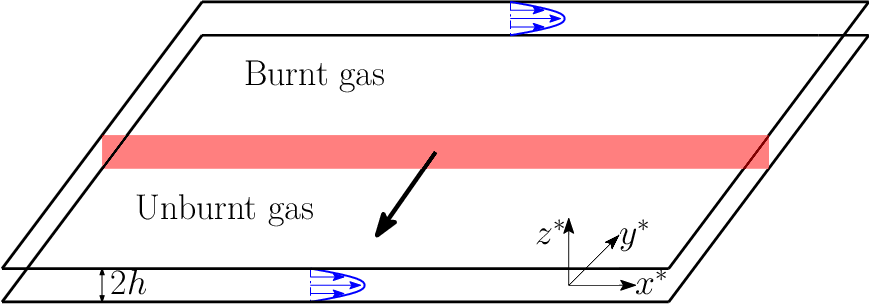}\vspace{0.1cm}
\caption{Schematic illustration of a planar premixed flame propagating in the presence of a shear flow in a narrow channel. The top subfigure corresponds to flame propagation in the streamwise $x^*$-direction and the bottom subfigure to propagation in the spanwise $y^*$-direction.} 
\label{fig:setup}
\end{figure}

It is worth pointing out that the shear flow induces shear-enhanced diffusion (Taylor dispersion) in the flow $x^*$-direction which can  most strongly influence the burning speed of flames propagating in this direction \cite{pearce2014taylor, daou2018taylor}, such as in the case shown in  Fig.~\ref{fig:setup}(a).  Since the enhancement of diffusion does not occur in the spanwise direction, the $y^*$-direction in Fig.~\ref{fig:setup},
 the diffusion is effectively anisotropic.  On account of the anisotropy,    the flame structure and instabilities are expected to be  significantly affected by the direction of flame propagation. This is particularly true when considering the direction of the planar flame whose stability is to be investigated;   
 two important cases, shown in Fig.~\ref{fig:setup}, are considered in this paper.  These  correspond  respectively to flames  propagating in the streamwise   $x^*$-direction,   and  the spanwise $y^*$-direction.  The second case, where the shear flow is parallel to the flame,    is particularly relevant when
 addressing  flame instabilities arising in a Taylor--Couette burner~\cite{vaezi2000laminar};  in this burner, the flame propagates parallel to the burner axis and  the shear flow takes place in the azimuthal direction. 
 
 In addition to the anisotropy aspect aforementioned, Taylor dispersion also produces another intriguing effect related to the presence of a flow-dependent effective Lewis number in the streamwise direction, as reported in~\cite{daou2018taylor,linan2020taylor}. Interestingly, the effective Lewis number approaches the inverse of the molecular Lewis number when the  shear flow intensity becomes large, and this has surprising repercussions on the diffusive-thermal flame instabilities  as reported in~\cite{daou2021effect,daou2023flame,daou2023diffusive} for premixed flames and~\cite{rajamanickam2023stability,kelly2023influence} for diffusion flames. As we shall confirm below, similar repercussions of the effective Lewis number, coupled with the Darrieus--Landau instability, will appear in our study.


\section{General formulation}    
\label{sec:basic}

In order to investigate the effect of a shear flow on the Darrieus--Landau instability, we shall adopt a plane  Poiseuille flow in a Hele-Shaw channel, although another flow profile such as a Couette flow can also be used. 
The channel half-width is denoted by $h$, and  the velocity field and   pressure gradient  are given in a frame moving with the mean flow speed $U$ by
\begin{equation}
    \vect v^* = \frac{U}{2}\left(1-\frac{3z^{*2}}{h^2}\right)\vect e_{x^*},  \quad \frac{dp^*}{dx^*}= -\frac{3\mu U}{h^2} \nonumber \label{poi}
\end{equation}
where $\vect e_{x^*}$ is a unit vector in the $x^*$-direction, $\mu$   the dynamic viscosity (assumed constant) and $p^*$  a modified pressure  which includes the hydrostatic pressure and the term $\tfrac{\mu}{3}\nabla\cdot\vect v^*$. We shall   consider flame propagation  in the narrow-channel limit $h \ll \delta_L$, where $\delta_L$ is the laminar flame thickness.  The chemistry is modelled by an irreversible reaction with an Arrhenius reaction rate $\rho^2 B Y_F e^{-E/RT}$,
where $\rho$ is the density, $B$ is the pre-exponential factor, $Y_F$ the mass fraction of the fuel (assuming to be the deficient species), $E/R$ the activation temperature and $T$  the temperature. The adiabatic flame temperature is defined by $T_{ad}=T_u(1+q)$, where $q=QY_{F,u}/c_p T_u$ is the heat release parameter, $c_p$   the specific heat at constant pressure and $Q$ is the amount of heat released per unit mass of fuel.  Here and below, the subscripts $u$ and  $b$ denote properties in the unburnt  and burnt gas, respectively. For simplicity, we shall assume that the channel walls are adiabatic, and that  $\mu$ and $\rho D$ are constant,  where $D$ is the thermal diffusivity. 

Introduce the non-dimensional variables
\begin{align}
    &t= \frac{t^*D_u}{\delta_L^2}, \quad \vect x=\frac{(x^*,y^*)}{\delta_L},\quad z=\frac{z^*}{h},\quad \varrho = \frac{\rho}{\rho_u}, \nonumber\\ &\vect v= \frac{\vect v^*}{S_L}, \quad
    p = \frac{p^* h^2/D_u}{\rho_uD_u\Pra},\quad Y=\frac{Y_F}{Y_{F,u}},\quad \theta=\frac{T-T_u}{T_{ad}-T_u} \nonumber
\end{align}
where $\vect x=(x,y)$ is the two-dimensional position vector, $\delta_L=D_u/S_L$   the laminar flame thickness, $S_L=\sqrt{2\Lew\beta^{-2}BD_u(\rho_b/\rho_u)^2e^{-E/RT_{ad}}}$  the laminar burning speed for $\beta\gg 1$, $\beta=E(T_{ad}-T_u)/RT_{ad}^2$   the Zeldovich number and $\Pra$  the Prandtl number. The non-dimensional governing equations in the low Mach-number approximation can then be written as 
\begin{align}   
    \frac{\varrho}{\Pra}\frac{D\vect v}{Dt} = -\frac{\nabla p}{\Dam} + \nabla^2\vect v - \frac{\Ray}{\Dam}(1-\varrho)\vect g ,\label{continuity}\\
     \frac{1}{\varrho}\frac{D\varrho}{Dt} =-  \nabla\cdot\vect v, \quad \varrho =1 / (1+q\theta) ,\label{eqnst} \\
    \varrho\frac{D\theta}{Dt}  = \nabla^2\theta + \omega,\quad
    \varrho\frac{DY}{Dt}  = \frac{1}{\Lew}\nabla^2 Y - \omega,   \label{TY}  
\end{align}
where $D/Dt=\partial_t+\vect v\cdot\nabla$, $\nabla=(\partial_x,\partial_y,\partial_z/\sqrt{\Dam})$, $\vect g=g_x\vect e_x+g_y \vect e_y$ is  a unit vector  in the direction of  gravity (with $g_x^2+g_y^2=1$) and
\begin{equation}
    \omega = \frac{\beta^2}{2\Lew}(1+q)^2\varrho^2 Y\exp\left[\frac{\beta(\theta-1)}{1+q(\theta-1)/(1+q)}\right]. \nonumber
\end{equation}
Furthermore, $\Dam=(h^2/D_u)/(\delta_L^2/D_u)$ is the Damk\"{o}hler number, $\Pec=Uh/D_u$  the flow Peclet number and $\Ray=\rho_u gh^2\delta_L/\mu D_u$ the Rayleigh number. Note that $\ep\equiv \sqrt{\Dam}=h/\delta_L$ is simply the ratio of the channel half-width to the laminar flame thickness. 

\section{Formulation for narrow channels}  
In the narrow-channel limit $\ep \equiv \sqrt{\Dam}\ll 1$, the dependent variables are expanded as
\begin{align}
   \vect v &= \ep^{-1}\vect v_0 +  \vect v_1(\vect x,z,t)  + \cdots, \nonumber\\
   \nabla p &= \ep^{-1}\nabla p_0+\nabla P(\vect x,t)+ \cdots , \nonumber\\
   \theta &= \theta_0(\vect x,t) + \ep\theta_1(\vect x,z,t)+\cdots \nonumber,\nonumber \\
   Y &= Y_0(\vect x,t) + \ep Y_1(\vect x,z,t)+\cdots \nonumber,\nonumber \\
   \varrho &=\varrho_0(\vect x,t) + \ep \varrho_1(\vect x,z,t) + \cdots. \nonumber
\end{align}
Substituting these into Eqs.~\eqref{continuity}-\eqref{TY}, and following  the derivation presented in~\cite{rajamanickam2022effects}, two-dimensional equations can be obtained. These govern flame propagation in narrow adiabatic channels in a frame moving with the mean flow speed in the unburnt gas.   Skipping algebraic details, the main results are as follows.

To leading order, the flow field  corresponding to the imposed Poiseuille shear flow, is given by
\begin{equation}
    \vect v_0 = \frac{\Pec}{2}(1-3z^2)\vect e_x, \quad \nabla p_0 = -3\Pec\,\vect e_x. \nonumber
\end{equation}
At the first order, we obtain $\varrho_1= -\theta_1d\varrho_0/d\theta_0$ as in~\cite{rajamanickam2023thick} and 
\begin{align}
    \theta_1&=\theta_c(\vect x,t) + \frac{\Pec}{8}\varrho_0\pfr{\theta_0}{x}(2z^2-z^4) , \nonumber\\
    Y_1&=Y_c(\vect x,t) + \frac{\Pec\Lew}{8}\varrho_0\pfr{Y_0}{x}(2z^2-z^4) , \nonumber \\
     v_{1,\vect x}&=\frac{1}{2}\left[\nabla P + \Ray(1-\varrho_0)\vect g\right](z^2-1) + \frac{\Pec^2}{40\Pra}\pfr{\varrho_0}{x}(5z^4-2z^6-3)\vect e_x, \nonumber \\
     v_{1,z} &= \frac{\Pec}{2\varrho_0}\pfr{\varrho_0}{x}(z^3-z), \nonumber   
\end{align}
where equations governing $(\theta_0,Y_0)$ and $(\theta_c,Y_c)$ can be obtained as solvability conditions of the second-order and third-order equations. The effective mass flux $\varrho_0\vect u$ due to gas expansion can be defined as
\begin{equation}
    \varrho_0\vect u  = \int_{0}^{1} (\varrho_0 \vect v_1 + \varrho_1 \vect v_0)dz. \nonumber
\end{equation}
Since the $z$-component of $\vect u$ is  zero,  we may regard $\vect u=(u,v)$ as a two-dimensional vector field.

The equations satisfied by $\theta_0(\vect x,t)$ and $Y_0(\vect x,t)$ are obtained, as mentioned above,  from the solvability condition of the second-order equations. Dropping the subscript $0$ in $\theta_0$, $Y_0$ and $\varrho_0$, the final two-dimensional governing equations are found to be
\begin{align}
&\frac{1}{\varrho}\frac{\mathcal D\varrho}{\mathcal Dt} = - \nabla\cdot\vect u, \label{cont}\\
&\vect u = -\frac{\nabla P}{3} - \frac{\Ray}{3}(1-\varrho)\vect g - \chi \gamma\Pec^2\pfr{\varrho}{x}\vect e_x, \label{vel} \\
  & \varrho\frac{\mathcal D\theta}{\mathcal Dt}= \pfr{}{x}\left[(1+\gamma \Pec^2\varrho^2)\pfr{\theta}{x}\right] + \pfr{^2\theta}{y^2}+\omega,  \label{theta} \\
    &\varrho\frac{\mathcal DY}{\mathcal Dt}= \pfr{}{x}\left[\frac{1+\gamma \Pec^2\Lew^2\varrho^2}{\Lew}\pfr{Y}{x}\right] + \frac{1}{\Lew}\pfr{^2Y}{y^2}-\omega \label{fuel}  
\end{align}
with $\mathcal D/\mathcal Dt=\partial_t+\vect u\cdot\nabla$, $\chi=1+3/\Pra$ and
\begin{equation}
    \gamma=\int_{0}^{1}d\zeta\left[\int_{0}^{\zeta}dz\, \frac{v_{0,x}}{\Pec}\right]^2=\frac{2}{105}. \nonumber
\end{equation}
By combining~\eqref{cont} and~\eqref{theta} and using~\eqref{eqnst}, we obtain
\begin{equation}
    \frac{\nabla\cdot\vect u}{q} = \pfr{}{x}\left[(1+\gamma\Pec^2\varrho^2)\pfr{\theta}{x}\right] + \pfr{^2\theta}{y^2}+\omega. \label{div}
\end{equation}
The equation leads, when used with ~\eqref{vel}
and the relation  $d\varrho/d\theta=-q\varrho^2$,  to the Poisson equation 
\begin{align}
    -\frac{\nabla^2 P}{3q}  = \pfr{}{x}\left\{\left[1+\gamma\Pec^2\varrho^2(1-\chi)\right]\pfr{\theta}{x}\right\}  + \pfr{^2\theta}{y^2}
    +\omega+ \frac{\Ray}{3}\varrho^2\vect g\cdot\nabla\theta  \label{poisson}
\end{align}
for the pressure field. Equations~\eqref{cont}-\eqref{poisson} generalize  similar equations used in the earlier study~\cite{fernandez2019impact} to non-zero values of $\Pec$. Notably, these equations clearly demonstrate the presence of anisotropic diffusion, and they will be used to study the stability of the planar flames in the two cases depicted in Fig.~\ref{fig:setup}. Furthermore, as can be from from~\eqref{div} the expansion rate $\nabla\cdot\vect u$ of a fluid element is directly affected by the anisotropic heat conduction. \blue{It is also worth emphasising the presence in the modified Darcy equation~\eqref{vel} of an \textit{effective force} induced by the imposed shear flow, given by $-\chi\gamma\Pec^2\pfi{\rho}{x}\vect e_x$. This term describes a novel aspect associated with the coupling between thermal expansion and Taylor dispersion, which has not been reported before and which has relevance well beyond combustion in areas involving transport phenomena. Furthermore,} Eq.~\eqref{vel} implies that the vorticity is given by
\begin{equation} \label{vort}
   \nabla\times\vect u = \chi\Pec^2 \pfr{^2\varrho}{x\partial y}\vect e_z -\frac{\Ray}{3} \vect g\times \nabla\varrho.
\end{equation}
This is a new result which indicates that curved flames in a Hele-Shaw channel always experience vorticity in the presence of Taylor dispersion, that is whenever $\sqrt{\gamma} \Pec \neq 0$, \blue{even in the absence of viscosity variations as assumed for simplicity in this paper.}  \blue{Equation~\eqref{vort} also} implies that the vorticity field remains zero everywhere, except in (the preheat and reaction zones of) the flame where spatial density  variations occur. \blue{As can be seen from the two-dimensional equations~\eqref{cont}-\eqref{fuel}, the Peclet number enters only as $\Pec^2$, a behaviour that was originally pointed out by Zeldovich~\cite{zeldovich1937asymptotic} in a related heat-transfer problem and obtained in other combustion studies such as in~\cite{daou2001flame} in the case of narrow channels. Therefore whether the shear flow is in the positive or negative $x$-directions is unimportant  in the narrow-channel limit, under the simplifying constant-viscosity assumption adopted herein. This directional symmetry of the imposed flow is broken when next-order corrections to~\eqref{cont}-\eqref{fuel} are taken into account~\cite{rajamanickam2023thick} or also when viscosity variations are taken into account. The reader may also refer to the recent study~\cite{miroshnichenko2020hydrodynamic} on the flame hydrodynamic instabilities based on a different  two-dimensional (Euler--Darcy) model taking into account viscosity variations.}


\section{Stability of a planar flame  propagating in the streamwise direction\label{sec:stream}}

Consider a planar flame propagating steadily in the negative $x$-direction 
as depicted in Fig.~\ref{fig:setup}(a). 
Let $S$ be the (non-dimensional) flame burning speed, which is also its speed with respect to the mean flow. It is worth pointing out that the flame speed with respect to the walls is then  $-\Pec/\sqrt{\Dam}+S$ given that $\Pec/\sqrt{\Dam}$ is the non-dimensional mean flow speed (in the unburnt gas). Since Equations~\eqref{cont}-\eqref{poisson} are  already written in a frame moving with the mean flow, it is sufficient in order to make  the planar flame steady  to shift the $x$-coordinate by the amount $St$. Therefore, effecting the transformation $x\to x+St$, the  planar flame corresponds now to a steady solution  depending only on $x$. This steady solution is denoted using overbars and  is characterised by a  velocity and pressure field given by
\begin{align}
   & \overline\varrho(\overline u + S) = S, \quad \overline v = -\frac{1}{3}\Ray(1-\overline\varrho) g_y, \nonumber\\
     &\frac{1}{3}\frac{d \overline P}{dx} = S\frac{\overline\varrho-1}{\overline\varrho}- \chi \gamma \Pec^2 \frac{d\overline\varrho}{dx} - \frac{\Ray}{3}  g_x (1- \overline\varrho), \nonumber
\end{align} 
and temperature and mass-fraction fields governed by
\begin{align}
    S\frac{d\overline\theta}{dx} = &\frac{d}{dx}\left[(1+\gamma\Pec^2\overline\varrho^2)\frac{d\overline\theta}{dx}\right]+ \omega(\overline\theta,\overline Y), \nonumber \\
    S\frac{d\overline Y}{dx} = &\frac{1}{\Lew}\frac{d}{dx}\left[(1+\gamma\Pec^2\Lew^2\overline\varrho^2)\frac{d\overline Y}{dx}\right]- \omega(\overline\theta,\overline Y) ,\nonumber
\end{align}
subject to the boundary conditions $\overline\theta(-\infty)=\overline\theta(+\infty)-1=0$ and $\overline Y(-\infty)-1=\overline Y(+\infty)=0$. 

\begin{figure}[h!]
\centering
\includegraphics[scale=0.45]{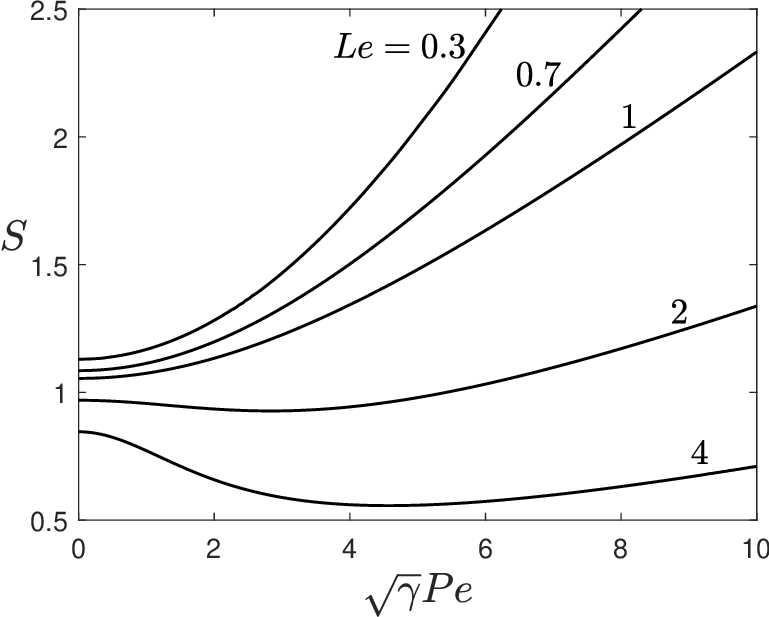}
\caption{The burning speed $S$ versus the Peclet number $\Pec$, computed for $\beta=10$, $q=5$, and selected values of $\Lew$, in the case of streamwise propagation.} \label{fig:base}
\end{figure}

\begin{figure*}[h!]
\centering
\includegraphics[scale=0.45]{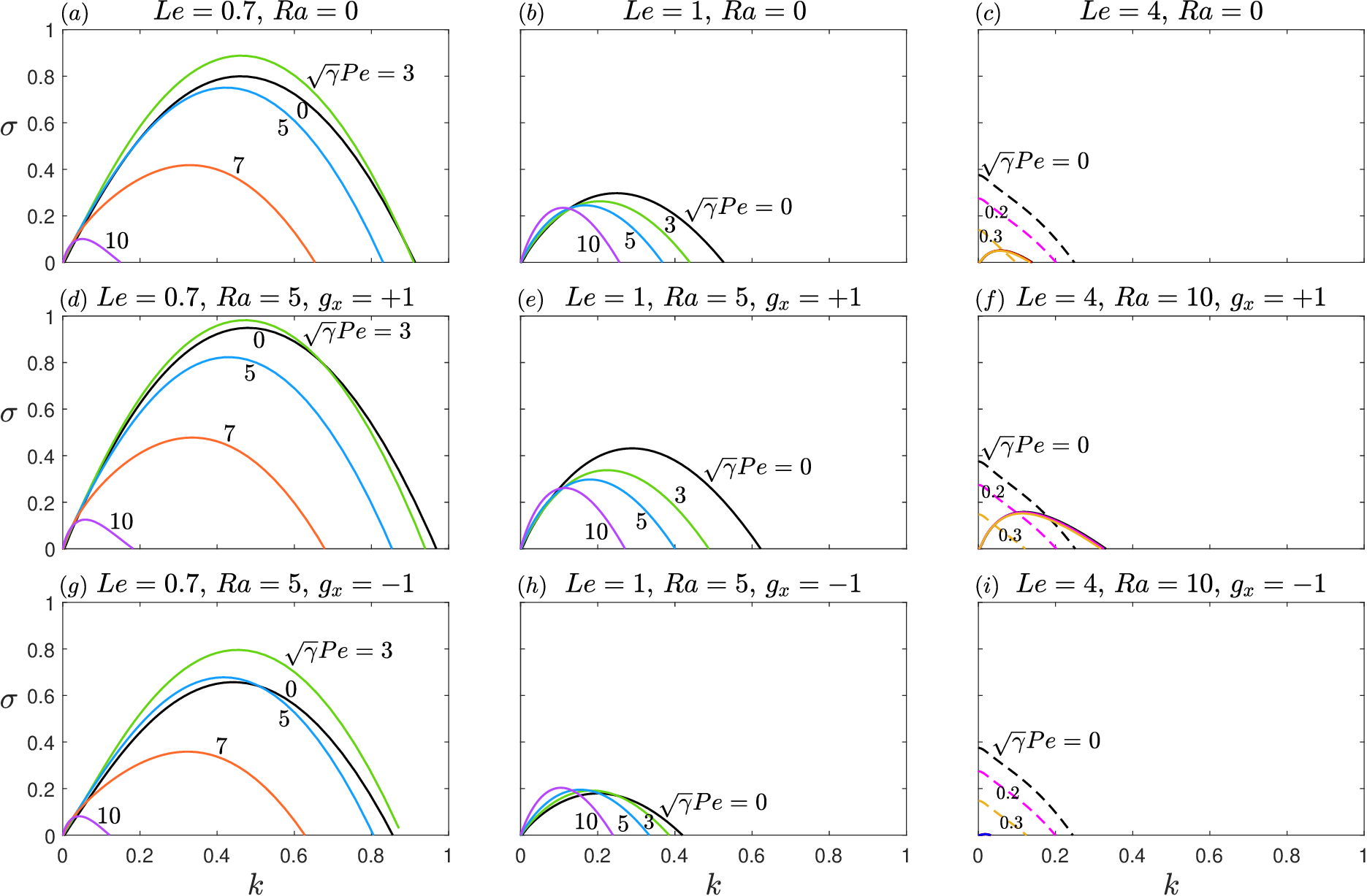}
\caption{Growth rate $\sigma$ versus the spanwise wavenumber $k$ for selected  values of $\Lew$, $\Pec$, and $\Ray$, in the case of streamwise propagation. 
The left, middle and  right columns    correspond respectively to $\Lew=0.7$, $\Lew=1$ and $\Lew=4$. The top, middle  and   bottom rows correspond respectively  to $\Ray=0$ (horizontal channels), $\Ray \neq 0$  with $g_x=+1$ (gravity vector pointing towards burnt gas) and $\Ray \neq 0$  with $g_x=-1$ (gravity vector pointing towards unburnt gas). Dashed lines in subplots $(c)$, $(f)$ and $(i)$ correspond to $\mathrm{Re}(\sigma)$ rather than $\sigma$ which is  complex with $\mathrm{Im}(\sigma)\neq 0$, indicating the presence of unstable oscillatory modes in these cases. In each of these subfigures, the solid lines represent unstable modes with real $\sigma$ for the three values of $\Pec$ indicated which are indiscernible (for such small values).} \label{fig:stream}
\end{figure*}

Note that the  burning speed $S$  depends on the imposed shear flow  whenever $\Pec\neq 0$, due to shear-enhanced diffusion.  In the asymptotic limit $\beta\gg 1$,   $S=(1+\gamma\Pec^2\overline\varrho_b^2)/\sqrt{1+\gamma\Pec^2\Lew^2\overline\varrho_b^2}$ as shown in~\cite{daou2018taylor,rajamanickam2023thick}.  Here,  we shall simply compute $S$ numerically for $\beta=10$ and $q=5$, and the results are shown in~Fig.~\ref{fig:base} for selected values of $\Lew$.

To investigate the stability of the planar flame described above, we introduce
\begin{equation}
\begin{bmatrix}
  \theta \\
  Y\\
 P
\end{bmatrix}=
    \begin{bmatrix}
\overline \theta(x) \\
\overline Y(x)\\
\overline P(x)
\end{bmatrix}+ \begin{bmatrix}
\hat \theta(x) \\
\hat Y(x) \\
\hat P(x)
\end{bmatrix}e^{iky+\sigma t}  \label{streamnormal}
\end{equation}
where the hat variables represent infinitesimal perturbations, $k$ is a real-valued spanwise wavenumber and $\sigma$ is, in general, a complex-valued  growth rate, which is to determined as an eigenvalue. The  perturbations or hat variables are governed by  linear equations which are given in the Appendix. These are solved numerically using COMSOL eigenvalue solver. All computations, here and below, are performed with the fixed values $\beta=10$, $q=5$ and $\Pra=0.72$  for selected values of $\Pec$, $\Lew$ and $\Ray$. \blue{Before presenting the results, it is worth noting that the corresponding results for $\Pec=0$ has been reported in the recent work~\cite{fernandez2019impact} based on a two-dimensional model that can be obtained from ours by setting $\Pec=0$ in~\eqref{cont}-\eqref{fuel}.}

The results for streamwise propagation are summarised in Fig.~\ref{fig:stream}. Shown is  the real growth rate $\sigma$  versus the spanwise wavenumber $k$ for selected values of $\Lew$, $\Ray$ and $\Pec$. The left, middle and  right columns correspond respectively to $\Lew=0.7$, $\Lew=1$ and $\Lew=4$. The top, middle  and   bottom rows correspond respectively  to $\Ray=0$ (horizontal channels), $\Ray \neq 0$  with $g_x=+1$ (gravity vector pointing towards burnt gas) and $\Ray \neq 0$  with $g_x=-1$ (gravity vector pointing towards unburnt gas). In each subfigure, several curves are plotted  for various values of $\Pec$, which are indicated. The following main observations can be inferred from Fig.~\ref{fig:stream}:
\begin{itemize}
    \item For $\Lew=1$ and $\Ray=0$, the effect of increasing $\Pec$ is to decrease  the maximum growth rate and to reduce the range of unstable wavenumbers. This  indicates an overall tendency  of the shear flow to impede the Darrieus-Landau instability. Note that the slope of the curves for small values of $k$ increases with increasing $\Pec$.  This is associated with the growth rate  being proportional to $S$ (for $k \ll 1$), and $S$ being an  increasing function of $\Pec$  as illustrated in Fig.~\ref{fig:base}.
    \item For  $\Ray\neq 0$, the Rayleigh--Taylor instability is seen to destabilize the flame when the gravity vector points towards to the  burnt gas (middle row) and to stabilise otherwise (bottom row), as expected.  It is worth noting that the effect of gravity on the dispersion curves  becomes weak for large values of $\Pec$; see the curves corresponding to $\sqrt{\gamma} \Pec=10$.
    \item For $\Lew=0.7$, the growth rates are larger compared to those of the unit Lewis number cases, presumably due  to such flames being prone to the cellular diffusive-thermal instability in addition to the  Darrieus--Landau instability. Note that the effect of an increase in $\Pec$ is stabilising for sufficiently large values of $\Pec$, but can be destabilising at smaller values as seen in the figure. 
    \item For $\Lew=4$, in addition to the Darrieus--Landau instability, whose dispersion curves are plotted as   solid lines, we also encounter the oscillatory diffusive-thermal instability, whose dispersion curves are plotted as dashed lines, indicating complex-valued growth rates. The latter instability is very sensitive to $\Pec$, and disappear with slight increases in $\Pec$,  as found in~\cite{daou2021effect}.  
\end{itemize}

\section{Stability of a planar flame  propagating in the spanwise direction\label{sec:span}} 

\begin{figure*}[h!]
\centering
\includegraphics[scale=0.45]{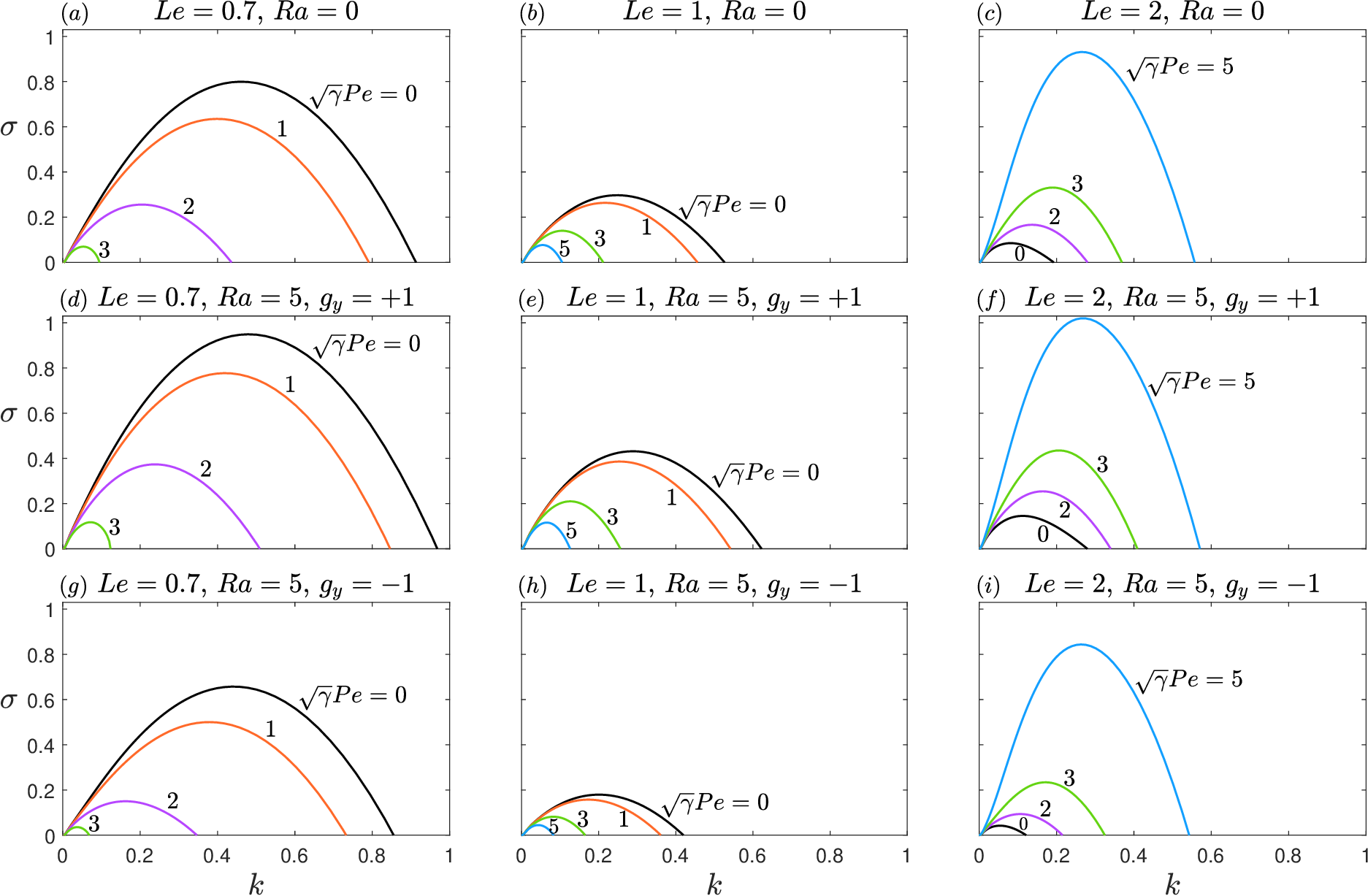}
\caption{Growth rate $\sigma$ versus the streamwise wavenumber $k$ for selected  values of $\Lew$, $\Pec$, and $\Ray$, in the case of spanwise propagation. 
The left, middle and  right columns    correspond respectively to $\Lew=0.7$, $\Lew=1$ and $\Lew=2$. The top, middle  and   bottom rows correspond respectively  to $\Ray=0$ (horizontal channels), $\Ray \neq 0$  with $g_y=+1$ (gravity vector pointing towards burnt gas) and $\Ray \neq 0$  with $g_y=-1$ (gravity vector pointing towards unburnt gas). } \label{fig:span}
\end{figure*}
 
We consider now the case depicted in Fig.~\ref{fig:setup}(b) where the unperturbed planar flame  propagates steadily in the negative $y$-direction with non-dimensional burning speed $S$. Unlike the case of 
Fig.~\ref{fig:setup}(a) addressed in the previous section, the planar flame and its burning speed are unaffected  by the shear flow. The corresponding velocity and   pressure field are given by
\begin{equation}
    \overline u = -\frac{1}{3}\Ray(1-\overline\varrho) g_x\quad \overline\varrho(\overline v + S) = S, \nonumber
\end{equation}
\begin{equation}
    \frac{1}{3}\frac{d \overline P}{dy} = S\frac{\overline\varrho-1}{\overline\varrho} - \frac{1}{3} \Ray g_y (1- \overline\varrho), \nonumber
\end{equation}
and the temperature and mass-fraction fields are governed by
\begin{equation}
    S\frac{d\overline\theta}{dy} - \frac{d^2\overline\theta}{dy^2} =\frac{1}{\Lew}\frac{d^2\overline Y}{dy^2}-S\frac{d\overline Y}{dy}= \omega(\overline\theta,\overline Y) \nonumber
\end{equation}
subject to $\overline\theta(-\infty)=\overline\theta(+\infty)-1=0$ and $\overline Y(-\infty)-1=\overline Y(+\infty)=0$. The value of $S$ here is independent of $\Pec$ and is equal to one in the asymptotic limit $\beta \gg 1$.  For the finite value  $\beta=10$  and $q=5$ adopted, $S$ takes the numerical values $1.085$, $1.054$, and $0.969$ for $\Lew=0.7$, $1$ and $2$, respectively. 

The stability of the planar flame  is investigated now by writing
\begin{equation}
\begin{bmatrix}
  \theta \\
  Y\\
 P
\end{bmatrix}=
    \begin{bmatrix}
\overline \theta(y) \\
\overline Y(y) \\
\overline P(y)
\end{bmatrix}+ \begin{bmatrix}
\hat \theta(y) \\
\hat Y(y) \\
\hat P(y)
\end{bmatrix}e^{ikx+\sigma t}  \label{spannormal}
\end{equation}
where the hat variables represent infinitesimal perturbations, $k$ the  streamwise wavenumber and $\sigma$ the growth rate. The linear system of equations governing the hat variables is given in the Appendix.

The results for spanwise propagation are summarised in Fig.~\ref{fig:span}. Shown is  the real growth rate $\sigma$  versus the streamwise wavenumber $k$ for selected values of $\Lew$, $\Ray$ and $\Pec$. The left, middle and  right columns correspond respectively to $\Lew=0.7$, $\Lew=1$ and $\Lew=2$. The top, middle  and   bottom rows correspond respectively  to $\Ray=0$ (horizontal channels), $\Ray \neq 0$  with $g_y=+1$ (gravity vector pointing towards burnt gas) and $\Ray \neq 0$  with $g_y=-1$ (gravity vector pointing towards unburnt gas). In each subfigure, several curves are plotted  for various values of $\Pec$, which are indicated. The following main observations can be inferred from Fig.~\ref{fig:span}:
\begin{itemize}
    \item For both $\Lew=1$ and $\Lew=0.7$ cases, the flame instability is impeded by increasing the shear strength. For $\Lew=2$, on the other hand, the instability  is seen to be promoted by the shear flow.
    \item More generally, it is important to note that whereas the flame instability is promoted by a decrease in the value of $\Lew$ for small or zero values of $\Pec$,
    it is promoted by an increase in $\Lew$ for larger values of $\Pec$. This is an interesting result which has been also found and attributed to Taylor dispersion for constant density diffusive-thermally unstable flames~\cite{daou2023diffusive}.   
\end{itemize}

\section{Time-dependent numerical simulations} 
\begin{figure*}[h!]
\centering
\includegraphics[scale=0.45]{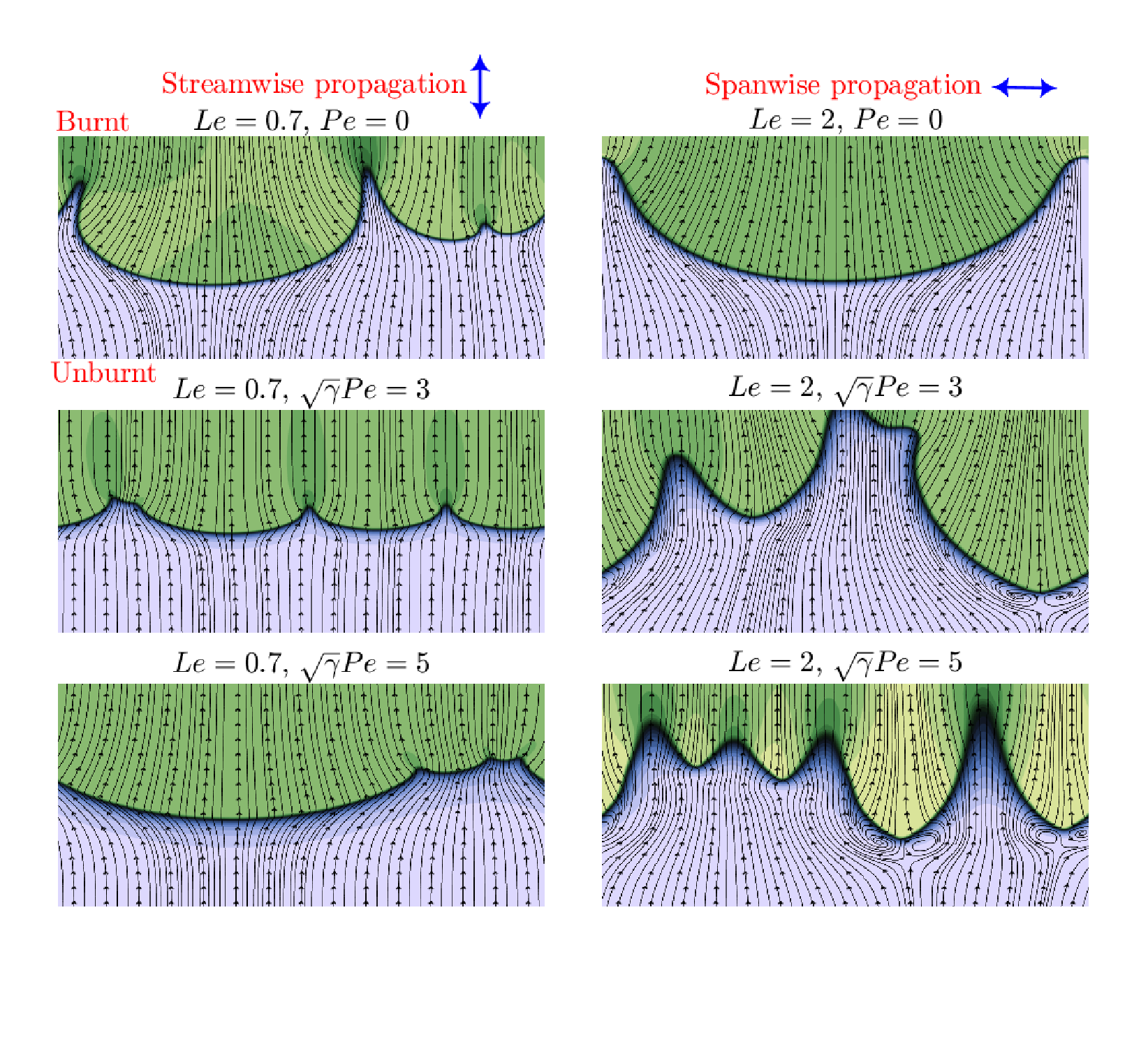} 
\caption{Instantaneous temperature fields and streamlines seen from the frame moving the flame, for $\Ray=0$ \blue{(no gravity)}, $\beta=10$ and $q=5$. The left column corresponds to streamwise flame propagation \blue{with $\Lew=0.7$} and the right column to spanwise flame propagation \blue{with $\Lew=2$}. \blue{The double-headed arrows indicate the direction of the imposed shear flow.} The horizontal and vertical extent of these figures are $100$ and $80$ \blue{times the laminar flame thickness}. \blue{In the case of spanwise propagation, in addition to moving with the flame along the vertical direction, the adopted frame also moves horizontally with the constant mean flow.}} \label{fig:react}
\end{figure*}

Time-dependent computations using Eqs.~\eqref{vel}-\eqref{fuel} and~\eqref{poisson} and an initial condition  corresponding to a planar flame solution \blue{(to which small amplitude random perturbations are added)} are performed using COMSOL Multiphysics, \blue{following the methodology described in~\cite{daou2023flame,rajamanickam2023thick}}. \blue{A nonuniform grid with typically 300 000 triangular elements is chosen
with local refinement around the reaction zone.} In \blue{the case of streamwise propagation}, a frame moving in the negative $x$-direction with the flame is adopted whose speed with respect to the unburnt gas is taken to be equal to the total instantaneous burning rate $ \frac{1}{L}\int \omega \,dxdy$ per unit transverse domain size  $L$. \blue{With respect to the channel walls, the adopted frame moves with the speed $-\Pec/\sqrt{\Dam}+ \frac{1}{L}\int \omega \,dxdy$, where $\Pec/\sqrt{\Dam}$ is the non-dimensional mean flow speed of the unburnt gas, which can take both positive and negative values. In the case of spanwise propagation, the adopted frame moves in the negative $x$-direction with the speed $-\Pec/\sqrt{\Dam}$ and in the negative $y$-direction with the speed $ \frac{1}{L}\int \omega \,dxdy$. Note that the governing  Eqs.~\eqref{vel}-\eqref{fuel} and~\eqref{poisson}, are already written in a frame moving with the mean flow in the $x$-direction and are independent of the sign of $\Pec$, as noted at the end of section~3. The solutions obtained and described below are  thus valid both for positive and negative values of the mean flow speed $\Pec/\sqrt{\Dam}$.} The boundary conditions \blue{in the frame adopted} correspond  to $\theta=Y-1=\nabla P=0$ far ahead of the flame  in the unburnt gas, and to  $\nabla\theta=Y=P=0$ far in the burnt gas. Periodic boundary conditions are imposed in the transverse direction. The domain extent in the direction of propagation is taken equal to $200$ and $L=100$ in the transverse direction.

Illustrative results for six  cases are presented in Fig.~\ref{fig:react}. Shown are instantaneous  temperature fields and streamlines in the frame moving with the flame adopted at selected times long after the instability has developed. \blue{The double-headed arrow is in the direction of the imposed shear flow and indicates the equivalence of two opposite parallel directions of the flow. As emphasised at the end of section~3, the later equivalence ceases to hold when the constant-viscosity assumption is relaxed (see, for instance~\cite{miroshnichenko2020hydrodynamic}) or in  wider channels (see, e.g.,~\cite{rajamanickam2023thick}).}

Several important observations can be inferred from the figure (and from   examination of the animations provided as supplementary material). For streamwise propagation (the subfigures in the left column), it is seen that an increase in $\Pec$ tends to flatten and thicken the flame and weaken its wrinkles. 
As for   spanwise propagation (the subfigures in the right column),
the opposite tendency is observed. It is worth noting that  the streamlines are deflected in an unusual fashion  for higher values of $\Pec$, with the  appearance of  a stagnation point  within the flame  preheat zone  accompanied by the presence of counter-rotating vortices; see middle and bottom subfigures in right column and compare with top subfigure. The presence of such vortices, and more generally the presence of vorticity within the flame and nowhere else around it, can be explained by the remarks following Eq.~\eqref{vort}.  Finally, note  the unusual presence of  structures with pointed leading edges close to  the counter-rotating vortices. The reader is referred to the supplementary materials for a better appreciation of the full time evolution of the unstable flames.

\section{Conclusions} 

In this paper, we have investigated the effect of a shear flow on the Darrieus--Landau instability in a Hele-Shaw channel. It is found that a strong shear flow  tends  to suppress the flame instability in the case of streamwise flame propagation. On the other hand, for flame propagation in the spanwise direction, the shear flow is found to have a destabilizing effect in $\Lew>1$ mixtures and a stabilizing effect in $\Lew<1$ mixtures. Furthermore,  for unstable flames propagating in the streamwise direction, the flow tends to mitigate flame wrinkling,  whereas the opposite tendency is encountered for flames propagating in the spanwise direction. An original result revealed by this study, is that curved flames in a Hele-Shaw channel always experience vorticity in the presence of Taylor dispersion, that is whenever $\sqrt{\gamma} \Pec \neq 0$, even in the absence of gravity \blue{and viscosity variations}.  

 \blue{One of the interesting findings correspond to the  peculiar  flame shapes encountered including cusps pointing towards to the unburnt gas and counter-rotating vortices in the case of spanwise flame propagation. This finding should stimulate future experimental investigations to observe such patterns, especially within the Taylor--Couette burner configuration~\cite{vaezi2000laminar}, where the direction of the shear flow is transverse to that of flame propagation. Finally, the analysis of the effect of Taylor dispersion on the flame hydrodynamic instabilities, explored here for a steady imposed flow, is worth extending to other shear flows; these include oscillatory flows~\cite{watson1983diffusion} and spatio-temporal periodic flows~\cite{rajamanickam2023effective}. Such extension would complement recent theoretical and experimental findings on flame instabilities in such oscillatory flows~\cite{radisson2022forcing}.}

\section*{Declaration of competing interest} 

The authors declare that they have no known competing financial interests or personal relationships that could have appeared to influence the work reported in this paper.

\section*{Acknowledgments} 

This work was supported by the UK EPSRC through grant EP/V004840/1.

\section*{Supplementary material} 

Animated time evolution of numerical simulations are included as video files.

\section*{Appendix} 

In this appendix, we write down the linear system  of equation corresponding to the linear stability analysis of the planar premixed flames propagating either in the streamwise direction as in Fig.~\ref{fig:setup}(a), or in the spanwise direction as in Fig.~\ref{fig:setup}(b). The system is  an eigen-boundary value problem for the perturbations   $\hat \theta$, $\hat Y$, and $\hat P$ introduced in Eqs.~\eqref{streamnormal} and \eqref{spannormal}.
For convenience, we also use the auxiliary variables  $\hat\varrho$, $\hat u$, $\hat v$, and $\hat\omega$, which are expressible in terms of $\hat \theta$, $\hat Y$, and $\hat P$.  For example, $\hat\varrho = -q\overline\varrho^2\hat\theta$ and 
\begin{align}
    \hat\omega= \frac{\beta^2}{2}(1+q)^2\exp\left[\frac{\beta(\overline\theta-1)}{1+q(\overline\theta-1)/(1+q)}\right]\left[\overline\varrho^2\hat Y + 2\overline\varrho\hat\varrho\overline Y + \frac{\overline\varrho^2\overline Y\beta\hat\theta}{[1+q(\overline\theta-1)/(1+q)]^2}\right]. \nonumber
\end{align}

For the case of streamwise propagation, we obtain the system of equations
\begin{align}
&3\hat u = - \hat P' - 3\chi \gamma\Pec^2 \hat\rho' + \Ray \hat \rho g_x, \nonumber\\
    &\sigma \overline \rho \hat \theta + S\hat\theta' + \overline\varrho(\hat u -qS\hat\theta) \overline\theta'  + ik\overline\rho\,\overline v \hat \theta = \hat\omega +\left[(1+\gamma\Pec^2\overline\rho^2)\hat\theta'+2\gamma\Pec^2\overline\rho\hat\rho \overline\theta'\right]'-k^2\hat\theta, \nonumber\\
&\sigma \overline \rho \hat Y + S\hat Y' + \overline\varrho(\hat u -qS\hat\theta) \overline Y'  + ik\overline\rho\,\overline v \hat Y = -\hat\omega + \frac{1}{\Lew}\left[(1+\gamma\Pec^2\Lew^2\overline\rho^2)\hat Y'+2\gamma\Pec^2\Lew^2\overline\rho\hat\rho \overline Y'\right]'-\frac{k^2\hat Y}{\Lew}, \nonumber\\
     &\frac{k^2\hat P-\hat P''}{3q} = \left\{\left[1+\gamma\Pec^2\overline\rho^2(1-\chi)\right]\hat\theta'\right\}' -k^2\hat\theta+  \hat\omega + 2\gamma\Pec^2(1-\chi)\left(\overline\rho\hat\rho \overline\theta'\right)' -  \frac{\Ray}{3q} (g_x \hat\varrho'+ikg_y\hat\rho)\nonumber  
\end{align}
subject to $\hat \theta=\hat Y=\hat P'=0$ as $x\to\pm\infty$. Here and below prime  denotes differentiation of a function   with respect to its argument. It can be checked that these equations reduce to those reported in~\cite{fernandez2019impact} when $\Pec=0$ and $g_y=0$. 

For the case of spanwise propagation, we obtain the system of equations 
\begin{align}
&3\hat v = - \hat P'  + \Ray \hat \rho g_y, \nonumber\\
    &\sigma \overline \rho \hat \theta + S \hat\theta' + \overline\varrho(\hat v -qS\hat\theta) \overline\theta'  + ik\overline\rho\,\overline u \hat \theta = \hat\theta''-k^2(1+\gamma\Pec^2\overline\rho^2)\hat\theta+\hat\omega, \nonumber\\
    &\sigma \overline \rho \hat Y + S \hat Y' + \overline\varrho(\hat v -qS\hat\theta) \overline Y'  + ik\overline\rho\,\overline u \hat Y = \frac{\hat Y''}{\Lew}-\frac{k^2}{\Lew}(1+\gamma\Pec^2\Lew^2\overline\rho^2)\hat Y-\hat\omega, \nonumber\\
     &(k^2\hat-\hat P'')/(3q)=\hat\theta'' - k^2[1+\gamma\Pec^2\overline\rho^2(1-\chi)]\hat\theta+\hat\omega   -  \Ray(g_y\hat\varrho'+ikg_x\hat\varrho)/(3q)\nonumber   
\end{align}
 subject to $\hat \theta=\hat Y=\hat P'=0$ as $y\to\pm\infty$.

\bibliographystyle{elsarticle-num}

\bibliography{elsarticle-template}

\begin{thebibliography}{10}
\expandafter\ifx\csname url\endcsname\relax
  \def\url#1{\texttt{#1}}\fi
\expandafter\ifx\csname urlprefix\endcsname\relax\def\urlprefix{URL }\fi
\expandafter\ifx\csname href\endcsname\relax
  \def\href#1#2{#2} \def\path#1{#1}\fi

\bibitem{veiga2020unexpected}
F.~Veiga-L{\'o}pez, M.~Kuznetsov, D.~Mart{\'\i}nez-Ruiz, E.~Fern{\'a}ndez-Tarrazo, J.~Grune, M.~S{\'a}nchez-Sanz, Unexpected propagation of ultra-lean hydrogen flames in narrow gaps, Phys. Rev. Lett. 124~(17) (2020) 174501.

\bibitem{gu2021propagation}
G.~Gu, J.~Huang, W.~Han, C.~Wang, Propagation of hydrogen--oxygen flames in {H}ele-{S}haw cells, Int. J. Hydrog. Energy 46~(21) (2021) 12009--12015.

\bibitem{yanez2022velocity}
J.~Yanez, M.~Kuznetsov, F.~Veiga-L{\'o}pez, On the velocity, size, and temperature of gaseous dendritic flames, Phys. Fluids 34~(11).

\bibitem{al2019darrieus}
E.~Al~Sarraf, C.~Almarcha, J.~Quinard, B.~Radisson, B.~Denet, P.~Garcia-Ybarra, Darrieus--{L}andau instability and markstein numbers of premixed flames in a {H}ele-{S}haw cell, Proc. Combust. Inst. 37~(2) (2019) 1783--1789.

\bibitem{veiga2019experimental}
F.~Veiga-L{\'o}pez, D.~Mart{\'\i}nez-Ruiz, E.~Fern{\'a}ndez-Tarrazo, M.~S{\'a}nchez-Sanz, Experimental analysis of oscillatory premixed flames in a {H}ele-{S}haw cell propagating towards a closed end, Combust. Flame 201 (2019) 1--11.

\bibitem{tayyab2020experimental}
M.~Tayyab, B.~Radisson, C.~Almarcha, B.~Denet, P.~Boivin, Experimental and numerical lattice-{B}oltzmann investigation of the {D}arrieus--{L}andau instability, Combust. Flame 221 (2020) 103--109.

\bibitem{han2021effect}
Y.~Han, M.~Modestov, D.~M. Valiev, Effect of momentum and heat losses on the hydrodynamic instability of a premixed equidiffusive flame in a {H}ele-{S}haw cell, Phys. Fluids 33~(10).

\bibitem{dominguez2023stable}
A.~Dom{\'\i}nguez-Gonz{\'a}lez, D.~Mart{\'\i}nez-Ruiz, M.~S{\'a}nchez-Sanz, Stable circular and double-cell lean hydrogen-air premixed flames in quasi two-dimensional channels, Proc. Combust. Inst. 39~(2) (2023) 1731--1741.

\bibitem{fernandez2019impact}
D.~Fern{\'a}ndez-Galisteo, V.~N. Kurdyumov, Impact of the gravity field on stability of premixed flames propagating between two closely spaced parallel plates, Proc. Combust. Inst. 37~(2) (2019) 1937--1943.

\bibitem{daou2021effect}
J.~Daou, Effect of {T}aylor dispersion on the thermo-diffusive instabilities of flames in a {H}ele-{S}haw burner, Combust. Theory Model. 25~(4) (2021) 765--783.

\bibitem{daou2023flame}
J.~Daou, A.~Kelly, J.~Landel, Flame stability under flow-induced anisotropic diffusion and heat loss, Combust. Flame 248 (2023) 112588.

\bibitem{daou2023diffusive}
J.~Daou, P.~Rajamanickam, Diffusive-thermal instabilities of a planar premixed flame aligned with a shear flow, Combust. Theory Model. 28~(1) (2024) 20--35.

\bibitem{pearce2014taylor}
P.~Pearce, J.~Daou, Taylor dispersion and thermal expansion effects on flame propagation in a narrow channel, J. Fluid Mech. 754 (2014) 161--183.

\bibitem{daou2018taylor}
J.~Daou, P.~Pearce, F.~Al-Malki, Taylor dispersion in premixed combustion: Questions from turbulent combustion answered for laminar flames, Phys. Rev. Fluids 3~(2) (2018) 023201.

\bibitem{vaezi2000laminar}
V.~Vaezi, R.~C. Aldredge, Laminar-flame instabilities in a {T}aylor--{C}ouette combustor, Combust. Flame 121~(1-2) (2000) 356--366.

\bibitem{linan2020taylor}
A.~Li{\~n}{\'a}n, P.~Rajamanickam, A.~D. Weiss, A.~L. S{\'a}nchez, Taylor-diffusion-controlled combustion in ducts, Combust. Theory Model. 24~(6) (2020) 1054--1069.

\bibitem{rajamanickam2023stability}
P.~Rajamanickam, A.~Kelly, J.~Daou, Stability of diffusion flames under shear flow: Taylor dispersion and the formation of flame streets, Combust. Flame 257 (2023) 113003.

\bibitem{kelly2023influence}
A.~Kelly, P.~Rajamanickam, J.~Daou, J.~Landel, Influence of heat loss on the stability of diffusion flames in a narrow-channel shear flow, Combust. Sci. Technol.

\bibitem{rajamanickam2022effects}
P.~Rajamanickam, A.~D. Weiss, Effects of thermal expansion on {T}aylor dispersion-controlled diffusion flames, Combust. Theory Model. 26~(1) (2022) 50--66.

\bibitem{rajamanickam2023thick}
P.~Rajamanickam, J.~Daou, A thick reaction zone model for premixed flames in two-dimensional channels, Combust. Theory Model. 27~(4) (2023) 487--507.

\bibitem{zeldovich1937asymptotic}
Y.~B. Zeldovich, The asymptotic law of heat transfer at small velocities in the finite domain problem, Zh. Eksp. Teoret. Fiz 7~(12) (1937) 1466--1468.

\bibitem{daou2001flame}
J.~Daou, M.~Matalon, Flame propagation in poiseuille flow under adiabatic conditions, Combust. Flame 124~(3) (2001) 337--349.

\bibitem{miroshnichenko2020hydrodynamic}
T.~Miroshnichenko, V.~Gubernov, S.~Minaev, Hydrodynamic instability of premixed flame propagating in narrow planar channel in the presence of gas flow, Combust. Theory Model. 24~(2) (2020) 362--375.

\bibitem{watson1983diffusion}
E.~Watson, Diffusion in oscillatory pipe flow, J. Fluid Mech. 133 (1983) 233--244.

\bibitem{rajamanickam2023effective}
P.~Rajamanickam, J.~Daou, Effective {L}ewis number and burning speed for flames propagating in small-scale spatio-temporal periodic flows, Combust. Flame 258 (2023) 113077.

\bibitem{radisson2022forcing}
B.~Radisson, B.~Denet, C.~Almarcha, Forcing of a flame by a periodic flow in a {H}ele-{S}haw burner, Phys. Rev. Fluids 7~(5) (2022) 053201.

\end{thebibliography}

\end{document}